\documentclass{iau}
\usepackage{graphicx}
\usepackage[authoryear]{natbib}
\usepackage{aas_macros}

\title[Tidal shells as clocks of galaxy mergers] 
{Tidal shells as clocks of galaxy mergers and tests of cosmological models}

\author[
Ebrov{\'a}, B{\'i}lek, Sola, Duc, \& Lalovi{\'c}
]   
{Ivana Ebrov{\'a}\,$^{1,*}$,
            Michal B{\'i}lek\,$^{2}$,
            Elisabeth Sola\,$^{3}$, 
            Pierre-Alain Duc\,$^{4}$ ,
            Ana Lalovi{\'c}\,$^{2}$
}

\affiliation{
$^1$FZU -- Institute of Physics of the Czech Academy of Sciences, Na Slovance 1999/2, 182 00 Prague, Czechia; 
$^*$e-mail: ebrova.ivana@gmail.com \\
$^2$Astronomical Observatory, Volgina 7, 11060 Belgrade, Serbia \\
$^3$Institute of Astronomy, Madingley Rd, Cambridge CB3 0HA, UK \\ 
$^4$Universit{\'e} de Strasbourg, CNRS, Observatoire astronomique de Strasbourg (ObAS), UMR 7550, 67000 Strasbourg, France
}

\pubyear{2026}
\volume{403}  
\pagerange{1-10}
\jname{The Hidden Beauty of the Galactic Outskirts}
\editors{ }
\begin{document}

\maketitle

\begin{abstract}

Stellar shells\,--\,concentric arc-like tidal features formed predominantly in nearly radial minor or intermediate mergers\,--\,are observed in 10--20\% of early-type galaxies and a few percent of spirals. Their spatial and kinematic properties encode both the time since the merger and the host gravitational potential. Shells extend deep into galactic outskirts and can persist for several gigayears, offering a unique window into the late stages of galaxy mergers. 

The outermost observed shells are particularly useful, as they place strong constraints on the time and conditions of the merger. We show that at surface brightness limits of current and upcoming deep imaging surveys, cosmological simulations predict that a significant fraction of massive galaxies should host large ($>100$\,kpc) shells. Yet observationally, such extended shells remain extremely rare. This discrepancy may point to observational biases or missing elements in current cosmological simulations, offering a direct test of $\Lambda$CDM predictions at galactic scales.

As the direct comparison of simulated and observed tidal features seems problematic, we resort to comparison of merger histories. 
We developed and tested a method to estimate merger times from the spatial distribution of shells. We used this method to investigate a sample of nearby massive prolate rotators\,--\,early-type galaxies that appear to revolve around the major morphological axis. Cosmological simulations indicate their origin in recent significant mergers. Deep imaging reveals that all galaxies in our sample display signs of interactions, most often in the form of shells. This provides insight into the type and timing of mergers the galaxies underwent, which can be directly compared with simulations. We aim to expand this approach to large samples of shell galaxies identified in the upcoming large deep surveys. This will enable statistically robust constraints on the recent merger history of the Universe that, again, can be confronted with cosmological simulations. Moreover, it will provide a tool for studying the impact of mergers on various astrophysical phenomena such as star formation, triggering of active galactic nuclei, formation of kinematically peculiar galaxies, or the origin of transient events.

\end{abstract}

\firstsection 

\section{Introduction}

Galaxy mergers play a fundamental role in the evolution of massive galaxies.
They contribute to the buildup of stellar mass, modify galaxy morphology and kinematics, and leave behind long-lived tidal debris in the form of streams, tails, shells, and diffuse stellar halos.
Such structures provide valuable information about the assembly history of galaxies and offer a direct observational record of relatively recent merger events.

Among the most intriguing merger-related phenomena are prolate rotators\,--\,galaxies that rotate predominantly around their major morphological axis, also called ‘minor-axis rotation’ as the velocity gradient is observed along the minor morphological axis. 
This kinematic configuration is difficult to produce through secular evolution alone.
Numerical simulations suggest that prolate rotation is most commonly generated during specific types of galaxy mergers \citep[e.g.,][]{nb03,cox06,hof10,tsa17}. 
Merger connection is also suggested in cosmological simulations \citep{naa14,illprol,li18,bf19,sch18,sch20}. 
This makes prolate rotators particularly valuable laboratories for studying the consequences of merger-driven evolution.

In \cite{illprol}, we investigated prolate rotators in the Illustris hydrodynamical cosmological simulations \citep{vog14illpreintro,nel15illpub}.
We identified prolate rotators and analyzed their formation histories.
The simulations indicate that these systems are strongly linked to relatively recent significant mergers, typically occurring within the last $\sim6$\,Gyr, see also \cite{pr24} for an overview. 

Such mergers are expected to leave observable tidal debris in the outskirts of galaxies, including tidal tails, streams, shells, and asymmetric stellar halos.
Consistent with this expectation, in \cite{probs} we examined 19 observed prolate rotators with available deep optical imaging and found morphological signs of galaxy interaction in all cases.
This represents a statistically significant excess compared to a general sample of early-type galaxies from the MATLAS survey \citep{matlas20}, providing further support for a close connection between prolate rotation and recent merger activity.

Among the various forms of tidal debris, stellar shells are particularly valuable, as they not only indicate the occurrence of a merger, but also preserve information about its timing and dynamical evolution.
The spatial distribution of shells can be used to estimate the time elapsed since the merger and to probe the gravitational potential of the host galaxy.
This makes shell systems a promising bridge between observations and cosmological simulations, where merger histories are known directly.

\begin{figure}[]
\begin{center}
 \includegraphics[width=1\linewidth]{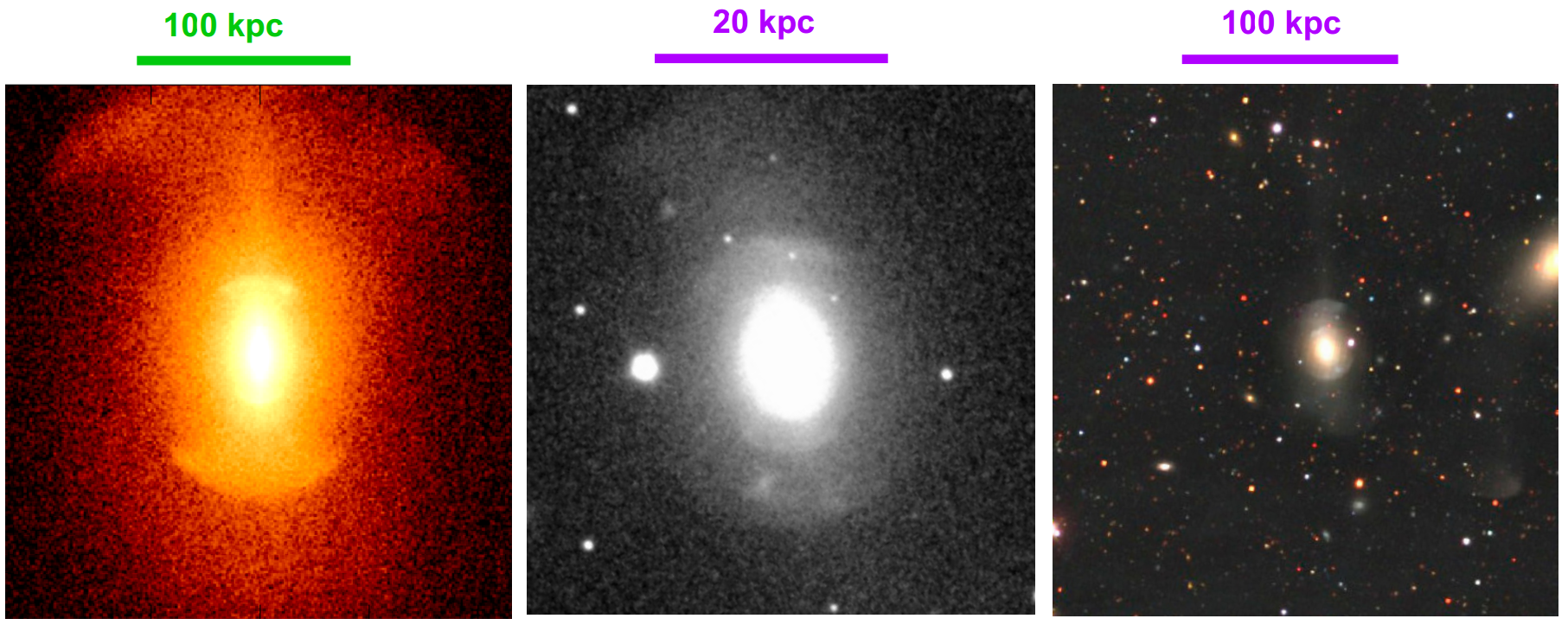} 
 \caption{
Comparism of simulated and observed shell -- left: a shell galaxy from Illustris simulations, middle: NGC\,7706 (SDSS), right: NGC\,7706 (DESI Legacy Imaging Surveys) in the same scale as the simulated galaxy. 
 }
   \label{fig:comp}
\end{center}
\end{figure}

\section{Shell sizes}

A natural next step is a more detailed comparison between simulated and observed prolate rotators. However, a direct comparison of their tidal features is not straightforward. As illustrated in Fig.\,\ref{fig:comp}, many simulated and observed systems exhibit remarkably similar shell morphologies, yet the physical scales of these structures differ substantially.

To quantify this discrepancy, we examined the sizes of the largest shells in the IllustrisTNG simulations. We generated $g$-band images for the 300 most massive galaxies in TNG100, adopting a surface brightness limit of 28.5\,mag\,arcsec$^{-2}$, comparable to that achieved by current ultra-deep imaging surveys such as MATLAS. 
We used the Jafar annotation tool \citep{matlas22} to document all visible shells, see an example in Fig.\,\ref{fig:comp}. 
Each galaxy was independently inspected by three team members. In total, we identified 65 shell galaxies containing 140 shells that were detected by at least two annotators.

\begin{figure}[]
\begin{center}
 \includegraphics[width=1\linewidth]{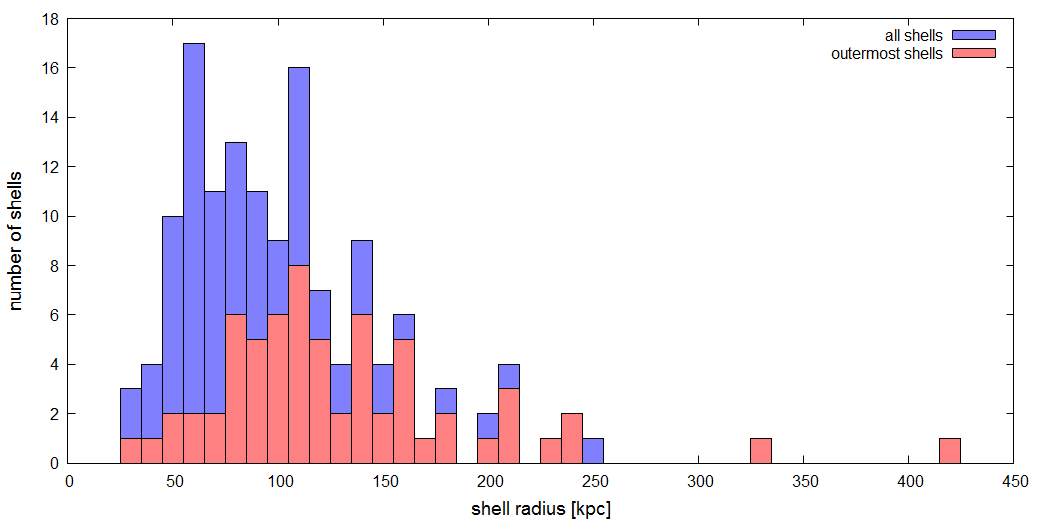} 
 \caption{
Histogram of shell radii detected in 300 most massive TNG100 galaxies.  
 }
   \label{fig:radii_hist}
\end{center}
\end{figure}

The distribution of shell radii is shown in Fig.\,\ref{fig:radii_hist}\,--\,the blue histogram represents all 140 shells, while the red histogram includes only the outermost shell in each galaxy.
A substantial fraction of the shells have radii exceeding 100\,kpc. 
This is surprising, as there are almost no observed shells of such a large size. 
\cite{matlas22} examined 179 early-type galaxies in the MATLAS survey and did not record any shell larger than 80\,kpc, while most shells in the survey have radii smaller than 40\,kpc.

One factor that may influence the visibility of outer shells in simulations is the mass resolution; however, as discussed in the next section, it does not appear to provide a straightforward solution to the problem. In cosmological simulations, a number of additional effects, often coupled to resolution, may also influence the appearance of shell systems.

In general, the radial distribution of shells is determined by the conditions of the merger that forms the system, with the most important parameters including the mass of the host galaxy, the impact velocity of the secondary galaxy, and the internal velocity dispersion of stars in the secondary. One possible explanation for the excess of very large shells in TNG100 is systematically high impact velocities: if, for example, the potential wells of host galaxies in cosmological simulations are deeper than those in the real Universe, this would lead to higher infall velocities of satellites and could consequently shift shell systems to larger radii.

In addition, other factors such as the internal structure of both the host and the secondary galaxy may play an important role by affecting the disruption process of the secondary system and the resulting phase-space distribution of debris. The origin of the excess of large shells in cosmological simulations therefore remains an open question.

\begin{figure}[]
\begin{center}
 \includegraphics[width=0.85\linewidth]{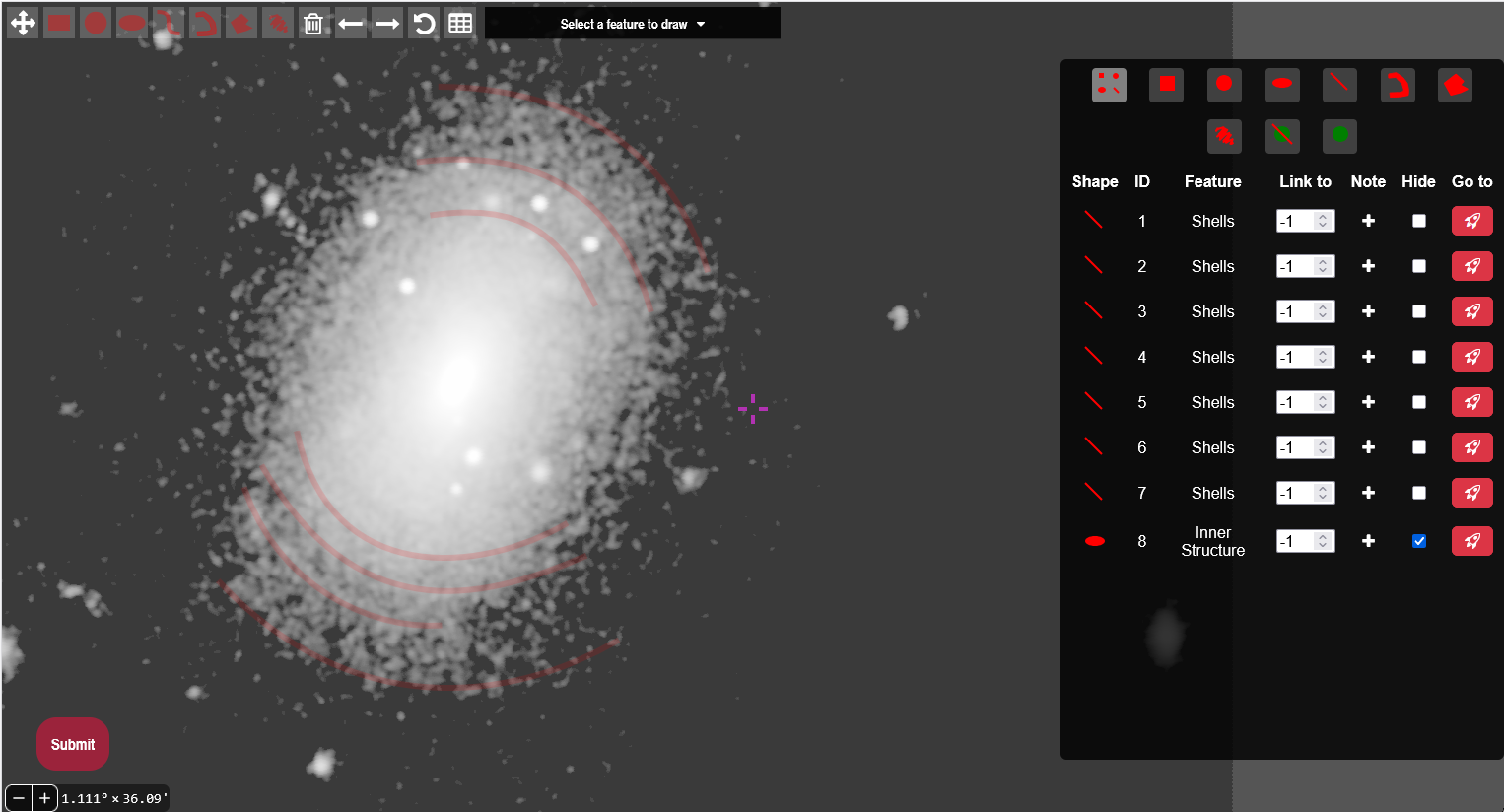} 
 \caption{
A TNG100 shell galaxy annotation the Jafar annotation tool.  
 }
   \label{fig:jafar}
\end{center}
\end{figure}

\section{Mass resolution effects}

The shell systems in TNG100 appear to differ from those observed in real galaxies in two important respects.
First, the simulation produces a large number of very extended shells.
Second, very inner shells, commonly observed in shell galaxies at galactocentric distances of only a few kiloparsecs, are largely absent.
The latter is not particularly surprising given the limited mass and spatial resolution of TNG100.
However, as discussed below, the difficulty of reproducing such inner shells may not be specific to cosmological simulations.

The lack of detected inner shells in our TNG100 sample is partly influenced by the design of our analysis.
The images were optimized for the detection of faint outer structures and are therefore less suitable for identifying small shells close to the galaxy center.
Some inner shells may thus be present in the simulations but remain undetected in our visual inspection.
Nevertheless, the abundance of large shells around TNG100 galaxies remains striking.

To explore whether these discrepancies could be related to mass resolution, we performed a simple numerical experiment using simulations of an isolated galaxy merger carried out with Gadget-2.
The detailed setup of the simulations is described in Appendix\,A of \cite{e20sg}.
We ran two simulations with identical initial conditions.
The only difference between them was the number of stellar particles representing the secondary galaxy and the corresponding gravitational softening length.
We then compared snapshots taken at the same time after the start of the simulation.

\begin{figure}[]
\begin{center}
 \includegraphics[width=1\linewidth]{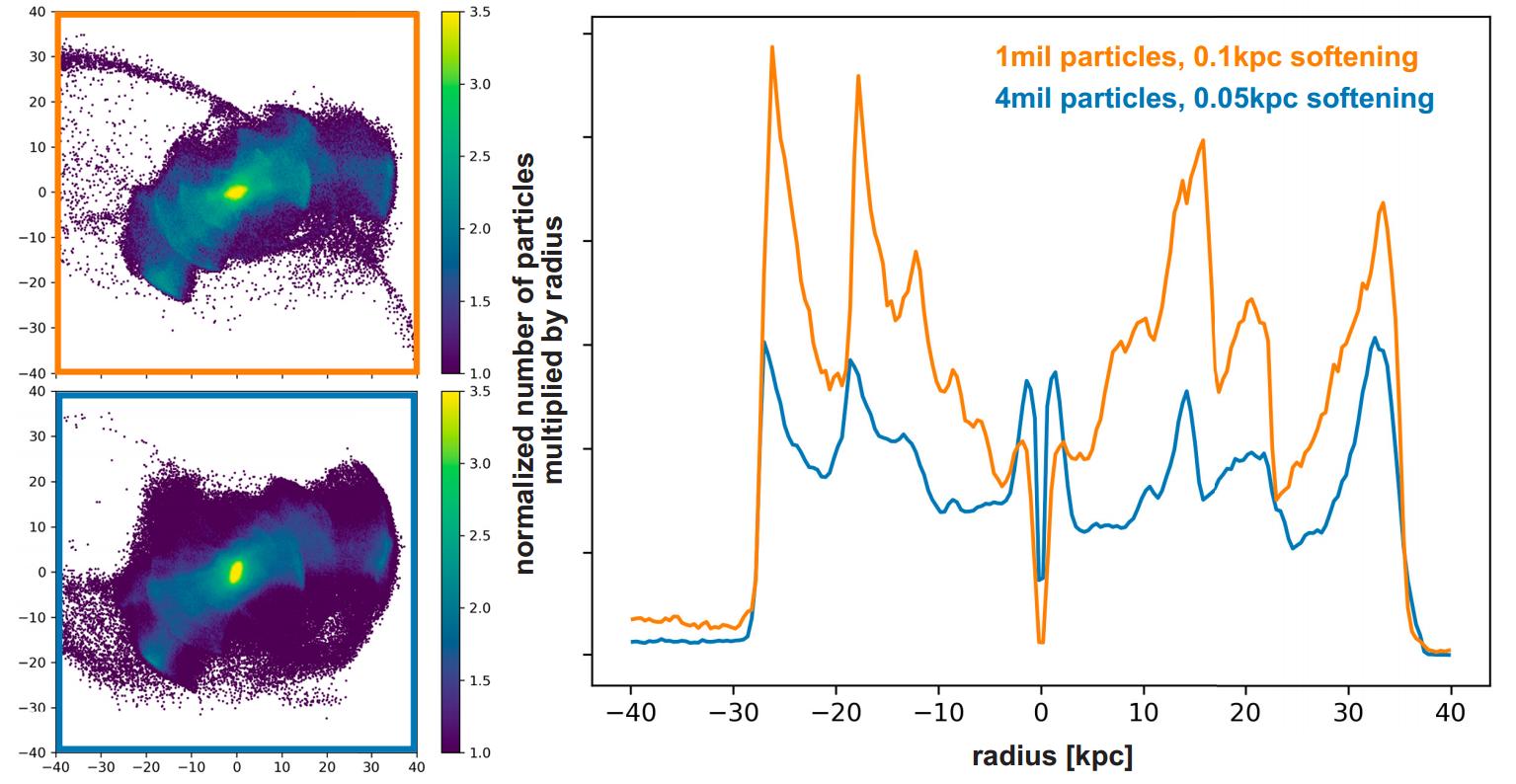} 
 \caption{
Surface brightness maps and radial histograms of the stellar component of the secondary galaxy in two merger simulations with different mass resolutions. 
 }
   \label{fig:res}
\end{center}
\end{figure}

Fig.\,\ref{fig:res} shows the surface-brightness maps and radial histograms of stellar particles belonging to the secondary galaxy.
The comparison reveals that increasing the mass resolution can reduce the surface brightness of the outer shells.
At the same time, it also decreases the visibility of shells at smaller galactocentric radii.
Some of the innermost shells even appear to be completely washed out in the higher-resolution simulation.

An important result is that the shell radii themselves remain unchanged.
The different resolutions affect the visibility of shells, but not the positions of those that remain detectable.

Although informative, this experiment is too limited to support general conclusions.
Based on this single comparison, the increase in mass resolution does not appear to provide a straightforward solution to the excess of large shells in TNG100.
While higher resolution may decrease the visibility of the outermost shells, it simultaneously reduces the visibility of the entire shell system.
If the outer shells become too faint to be detected, the inner shells are likely to become difficult to detect as well.

Furthermore, this experiment highlights a separate issue concerning the formation of inner shells.
Observed shell galaxies frequently contain shells at galactocentric distances of only a few kiloparsecs, and sometimes even less.
While the absence of such structures in TNG100 can plausibly be attributed to resolution limitations, they also appear to be difficult to reproduce in dedicated isolated merger simulations with substantially higher resolution.
In our experiment, increasing the mass resolution did not improve the situation, and in fact, made the innermost shells less prominent.
This suggests that the scarcity of very inner shells may represent a more general problem of current shell simulations rather than a limitation specific to TNG100 or cosmological simulations.

\section{Merger times from shell systems}

The direct comparison of tidal features in observed and simulated galaxies appears to be problematic.
As discussed in the previous section, shell systems in IllustrisTNG often extend to significantly larger radii than those observed in real galaxies.
This makes a one-to-one comparison of shell morphologies and scales difficult.
Instead, we focus on a quantity that can be inferred both from simulations and observations: the time elapsed since the merger. 

Galaxies containing multiple stellar shells are particularly well suited for this purpose.
The origin and evolution of shell systems were first described by \cite{q84} and subsequently studied by many authors \citep[e.g.][]{dc86,mk98,e12sg,sh13,guo24}. 
Their morphology suggests that they are produced by nearly radial mergers with unequal mass ratios.
In this scenario, stars originating from the less massive galaxy are released onto highly eccentric orbits.
Shells are observed near the apocenters of these trajectories, where stars spend a substantial fraction of their orbital periods.
Shortly after the merger, shells appear at low galactocentric radii.
As time progresses, successive shells move outward and the shell system becomes richer and more extended. 

The evolution of a shell system is primarily determined by the gravitational potential of the host galaxy.
For a given potential, it is possible to model the shell distribution expected at any time after the merger.
By comparing modeled and observed shell radii using the so-called ‘shell identification method’ \citep{bil13,bil14}, one can estimate the time elapsed since the stars were stripped from the secondary merger progenitor.
In many cases, stripping does not occur during a single encounter.
Instead, only a fraction of stars is released during the first close pericentric passage, while additional shell generation or generations are produced during subsequent passages of the surviving core through the host galaxy. 

The method has already been tested and applied in several studies.
In \cite{e20sg}, we validated the approach using numerical simulations and applied it to the shell galaxy NGC\,4993 in order to constrain the origin of the binary neutron-star merger responsible for the famous kilonova \citep{sss17a} and the GW170817 gravitational-wave event \citep{gw17}.
In \cite{bil22}, we further verified the method using simulations and combined it with multiple observational datasets to derive a detailed scenario of the origin of the spectacular shell system of NGC\,474. 

\begin{figure}[]
\begin{center}
 \includegraphics[width=1\linewidth]{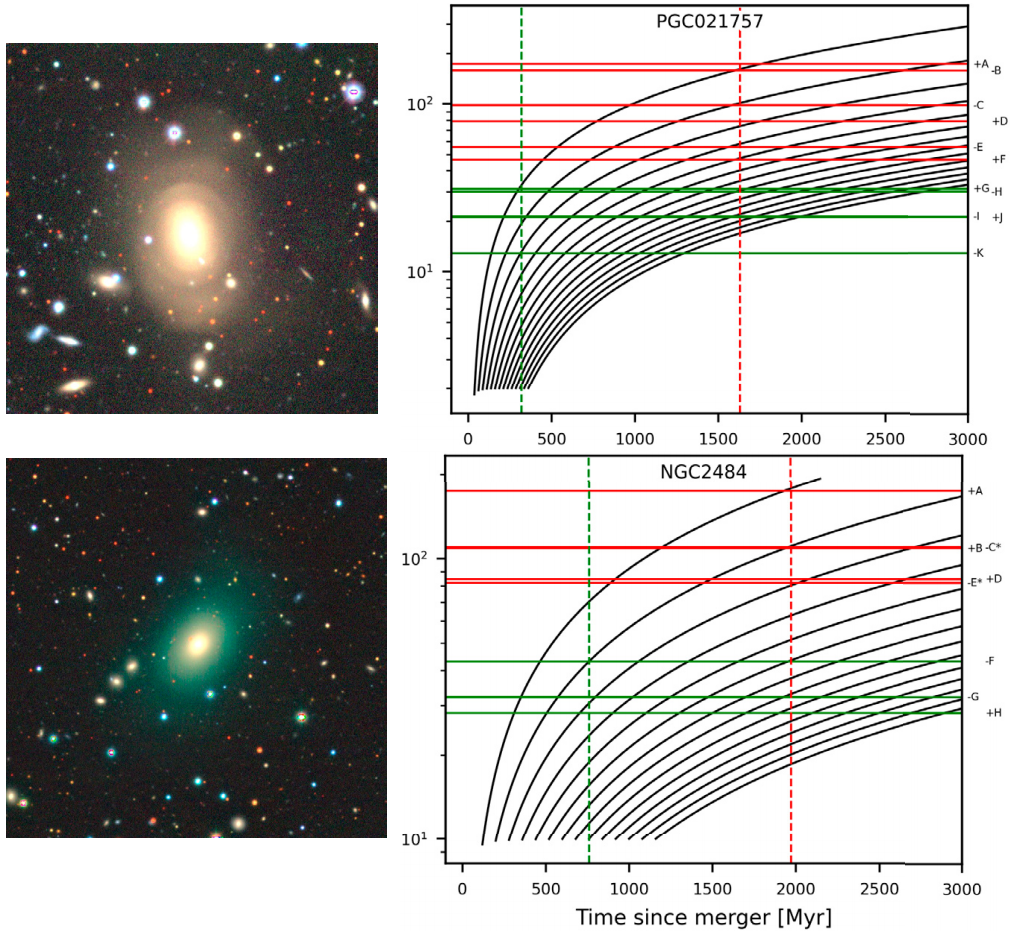} 
 \caption{
Two examples of shell galaxies with prolate rotation -- left: Legacy Surveys images; right: black curves show modelled evolution of shell radii, horizontal lines mark radii of measured shells, dashed vertical lines highlight the inferred age of each shell generation. 
 }
   \label{fig:prrolrot_shelliden}
\end{center}
\end{figure}

\begin{figure}[]
\begin{center}
 \includegraphics[width=1\linewidth]{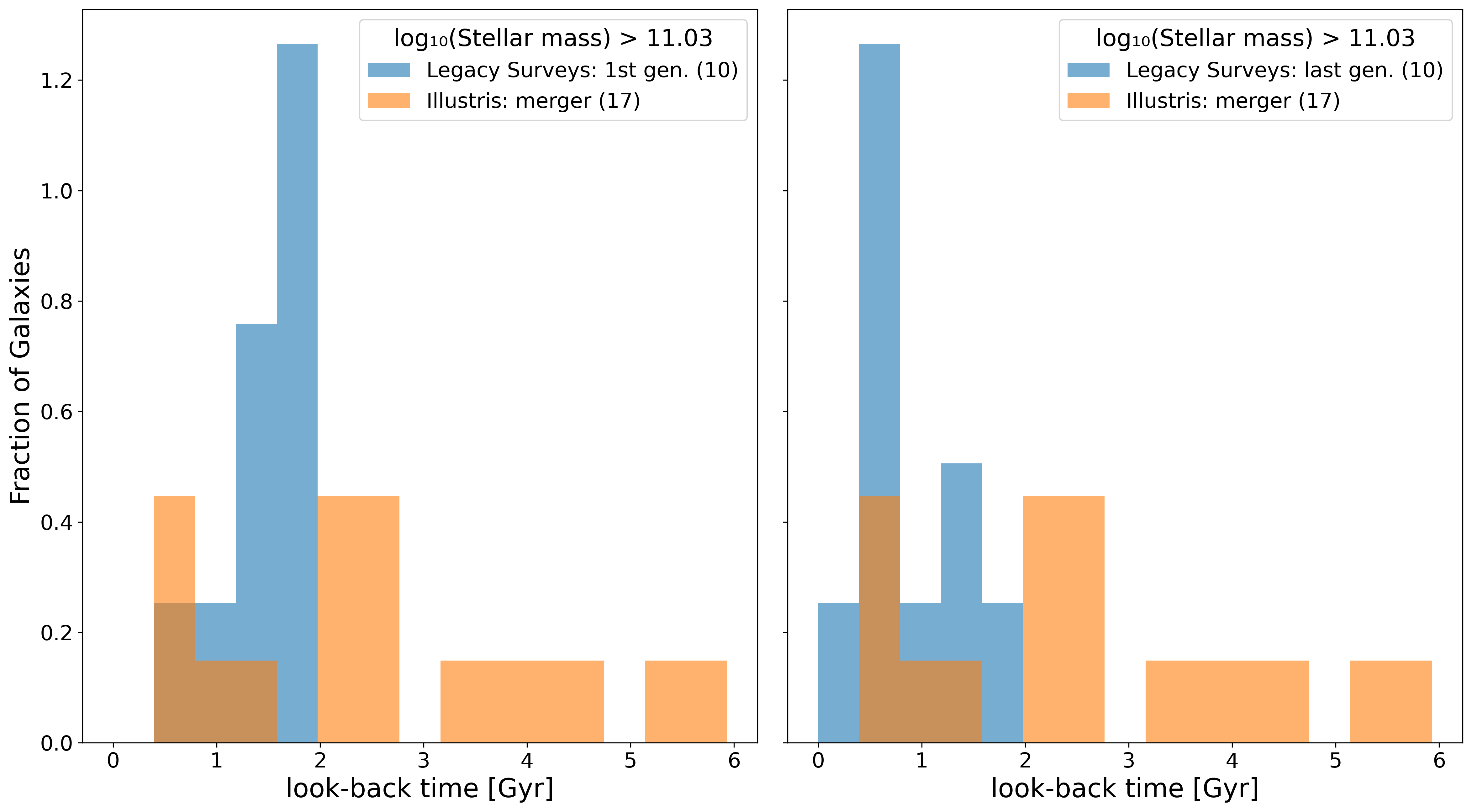} 
 \caption{
 Distribution of merger times for observed and simulated prolate rotators. Merger times of observed galaxies were estimated using the shell-identification method, while merger times of Illustris galaxies were obtained from merger trees. The left panel shows a comparison with the ages of the first identified shell generations, whereas the right panel shows a comparison with the ages of the last identified shell generations. Only galaxies with $\log(M_\star/M_\odot)>11.3$ are included. 
 }
   \label{fig:merger_times}
\end{center}
\end{figure}

\section{Shell-identification analysis of prolate rotators}

Here we are going to apply the method, for the first time, to a set of galaxies. 
Prolate rotators provide a nice showcase of an interesting kinematic phenomenon with a strong connection to galaxy mergers that can be studied in this way. 
From our original sample of 19 prolate rotators \citep{probs}, 11 are galaxies with multiple shell systems that have fairly deep images available in the DESI Legacy Imaging Surveys (hf. Legacy Surveys). 

For these 11 galaxies, we downloaded FITS images from Legacy Surveys DR9 and DR10 and identified all visible shells and other tidal features. 
To construct the gravitational potentials of the host galaxies, we used our recently developed \texttt{photomass\_ls.py} tool \cite{photomass}. 
The code derives stellar-mass distributions directly from Legacy Surveys imaging by converting observed surface brightness into stellar mass and constructing radial mass profiles suitable for dynamical modeling. The shell evolution was then computed using the Radial Acceleration Relation \citep{mcg16,lelli17,li18rsr}. Finally, we performed shell identification for all 11 galaxies. Two examples are shown in Fig.\,\ref{fig:prrolrot_shelliden} and full study will be published soon (Ebrov{\'a} et al., in prep). 

The resulting merger-time estimates were compared with those obtained for simulated prolate rotators from \cite{illprol}. For the Illustris sample, we defined the merger time as the last snapshot in which both merger progenitors were still listed as separate subhalos in the subhalo catalog. The precision of this estimate is limited by the temporal spacing of simulation snapshots, which is approximately 150\,Myr. 

To ensure a comparison between galaxies of similar masses, we considered only systems with stellar masses above $\log(M_\star/M_\odot)=11.3$. This selection yielded 10 observed galaxies and 17 simulated galaxies. The resulting merger-time distributions are shown in Fig.\,\ref{fig:merger_times}. Most observed galaxies required two, in some cases three, shell generations in order to reproduce the observed shell distribution. For this reason, the left panel compares the Illustris merger times with the ages of the first identified shell generations, while the right panel compares them with the ages of the last identified shell generations.

The oldest shell generations identified in our sample are approximately 2\,Gyr old.
This demonstrates that the shell-identification method can probe the recent merger history of shell galaxies over a substantial period of time.
However, it does not reproduce the full merger-time distribution found in Illustris, which contains a tail of mergers that occurred 3--6\,Gyr ago. 

Several explanations are possible.
The shell-identification method may become insensitive to older merger events.
Alternatively, shell systems produced by such mergers may be too faint to be detected in imaging of the depth provided by the Legacy Surveys.
This possibility could be tested with ongoing and upcoming surveys that reach significantly lower surface-brightness limits, such as those from the Vera C. Rubin Observatory and Euclid. 

Finally, we note that our observed sample contains only prolate rotators with multiple visible shell structures.
The simulated sample, on the other hand, includes all prolate rotators regardless of the presence of detectable tidal features.
This choice was motivated by the difficulties of directly comparing shell morphologies between observations and cosmological simulations, particularly in view of the differences in shell scales and the limited mass resolution of cosmological simulations.

\begin{figure}[]
\begin{center}
 \includegraphics[width=1\linewidth]{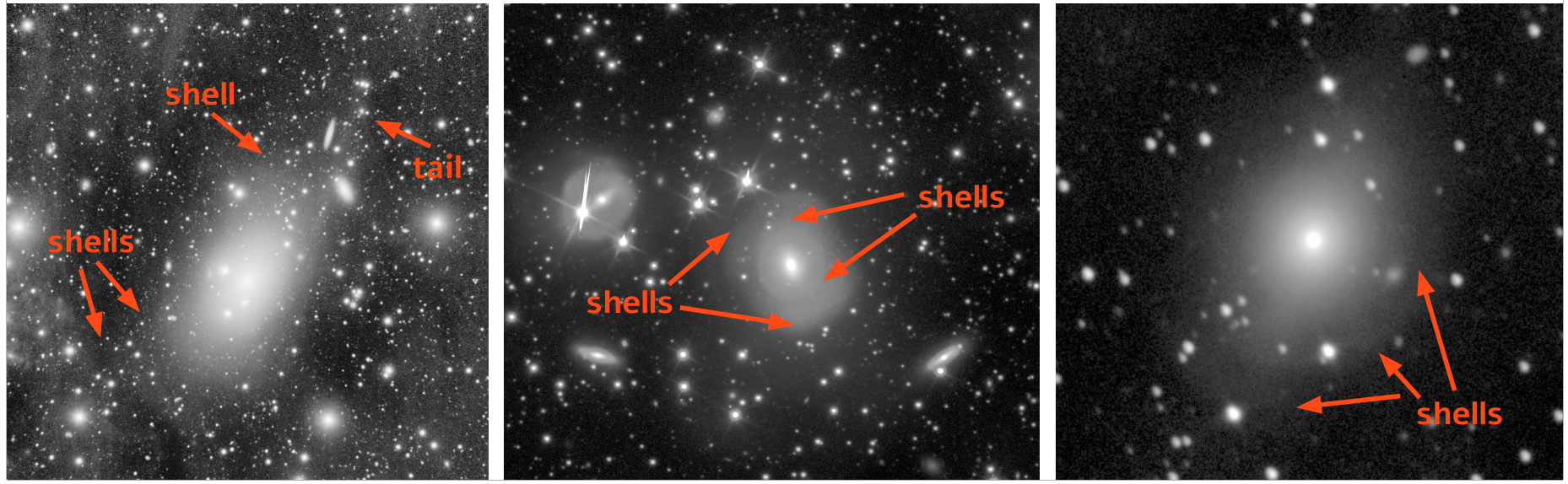} 
 \caption{
Examples of deep images of prolate rotators obtained with the Milankovi\'c telescope. All three galaxies exhibit clear signs of recent galaxy interactions, most prominently in the form of stellar shells. These observations are part of an ongoing effort to extend the sample of deeply imaged nearby massive prolate rotators beyond the coverage of the Legacy Surveys. 
 }
   \label{fig:milan}
\end{center}
\end{figure}

\section{Conclusions and future work}

In this work, we explored the connection between prolate rotation, stellar shells, and galaxy mergers using both observations and cosmological simulations.
Our previous studies demonstrated that prolate rotators are strongly linked to recent merger activity and that observed prolate rotators exhibit tidal features at a significantly higher rate than ordinary early-type galaxies.
Here, we extended this comparison by focusing on shell systems and the merger histories that they encode.

A direct comparison of shell morphologies between observations and the IllustrisTNG simulations revealed a significant discrepancy in scale.
While observed and simulated shell systems often display remarkably similar shapes, shells in TNG100 frequently reach much larger radii than those found around nearby galaxies.
A substantial fraction of simulated shells exceed 100\,kpc in radius, whereas shells larger than 80\,kpc are extremely rare in observational samples.
The origin of this discrepancy remains unclear.
Our exploratory resolution study suggests that it is unlikely to be explained solely by the mass resolution of the simulations, although other resolution-related effects may still play an indirect role.

Because shell sizes appear difficult to compare directly, we turned our attention to merger times.
Stellar shells provide a unique opportunity to reconstruct the recent merger history of galaxies through the shell-identification method.
Applying this technique to a sample of shell-bearing prolate rotators, we derived estimates of the times elapsed since the mergers that formed their shell systems and compared them with merger times measured directly in simulations.
The observed shell systems typically indicate mergers younger than approximately 2\,Gyr, whereas the simulated prolate rotators include a tail of population of mergers that occurred 3--6\,Gyr ago.
Whether this discrepancy reflects limitations of the shell-identification method, observational depth, or differences between simulated and real galaxies remains an open question.

The present study also highlights the broader potential of shell galaxies as probes of galaxy evolution.
Because shell systems preserve information about merger events over gigayear timescales, they offer a powerful means of reconstructing the recent assembly histories of galaxies, particularly early-type galaxies.
This opens a path toward studying the influence of mergers on a wide range of astrophysical phenomena, including the formation of kinematically peculiar galaxies, the triggering of nuclear activity, and the origin of transient events.

Our observational sample of prolate rotators with deep images is currently being expanded.
Using the Serbian 1.4m Milanković telescope, we are obtaining deep images of nearby massive prolate rotators that are not covered by the Legacy Surveys.
The goal is to extend the sample to encompass virtually all known nearby massive prolate rotators.
Preliminary results are highly encouraging.
Signs of galaxy interactions are present in all newly observed systems, and several galaxies exhibit extensive shell structures, including multiple shell generations.
Three examples from the Milanković observations are shown in Fig.\,\ref{fig:milan}.

The combination of increasingly deep imaging and shell-based dynamical modeling promises to substantially improve our understanding of the merger histories of prolate rotators and early-type galaxies in general.
Future surveys reaching even lower surface-brightness limits, together with larger observational samples and improved simulations, will enable more stringent tests of galaxy formation models and the role of mergers in shaping the present-day galaxy population.

\section*{Acknowledgements}
We acknowledges the support from the FORTE project, supported by the Ministry of Education, Youth, and Sports of the Czech Republic, co-funded by the European Union (No. CZ.02.01.01/00/22\_008/0004632). 
This research was supported by the Ministry of Science, Technological Development and Innovation of the Republic of Serbia (MSTDIRS) through contract no. 451-03-33/2026-03/200002 made with the Astronomical Observatory (Belgrade, Serbia).


\bibliographystyle{aasjournal}
\bibliography{prolrot.bib}
 

\end{document}